\documentstyle[11pt,newpasp,twoside,epsf]{article}

\reversemarginpar

\begin{document}
\title{3D Anelastic Simulations of Convection in Massive Stars}
\author{Michael Kuhlen, Stanford E. Woosley}
\affil{Department of Astronomy and Astrophysics, University of California at Santa Cruz, 1156 High Street, Santa Cruz, CA 95064}
\author{Gary A. Glatzmaier}
\affil{Department of Earth Science, University of California at Santa Cruz, 1156 High Street, Santa Cruz, CA 95064}
\begin{abstract}
After briefly describing the anelastic approximation and Glatzmaier's code, we present results from our preliminary studies of core convection during the hydrogen burning phase of a $15 M_\odot$ star, as well as our most recent results concerning convection in the oxygen shell of a $25 M_\odot$ star.
\end{abstract}
\vspace*{-0.25in}
\section{Introduction}
% - general bla bla
% - first 3D models
% - convection inherently 3D
% - issues in stellar evolution depend on convection
% - past work on subject
Two-dimensional studies of turbulent convection in the interiors of massive stars have been quite common (Deupree 2000; Asida \& Arnett 2000; Bazan \& Arnett 1998), but the problem is inherently three-dimensional and many aspects of the convective flow will not be reproduced correctly in two-dimensional models (Canuto 2000). Only recently has it become possible to begin tackling the problem of simulating convection in all its three-dimensional g(l)ory, and several major projects are under way to do just this (Dearborn et al. 2001; Brun 2002; Porter \& Woodward 2000). The price of this additional dimension is additional computational complexity, the need for faster computers and more data storage capabilities, not to mention difficulties in visualizing and analyzing the resulting data. In this paper we describe three-dimensional simulations based on anelastic hydrodynamics, which we hope capture the essential phenomena of the problems at hand, while allowing more rapid progress at a lower computational cost than comparable studies using fully compressible codes.

This paper is divided into three parts. After very briefly describing the anelastic approximation and Glatzmaier's implementation of it in Section 2, we present our hydrogen core convection model and results of this simulation in Section 3, and the oxygen shell convection model and results in Section 4.

In addition to the static plots shown here, please look at {\em http://www.supersci.org} for movies of these models.
%\vspace*{-0.1in}
\section{The Numerical Method}
%\vspace*{-0.1in}
% - briefly describe problem with Boussinesq and fully compressible
The simulations presented in this paper were performed using anelastic hydrodynamics. This approximation allows a background density stratification, but filters out short-period, acoustic oscillations, thereby allowing larger timesteps than fully compressible codes. We believe that these attributes make anelastic hydrodynamics an excellent method for studying stable convective regions in massive stars.
  \subsection{The Anelastic Approximation}
%   - filter out soundwaves
%   - can get larger timestep
%   - looking for ``average'' solution
%   - cannot model shocks or anything (super)sonic
%   - expand thermodynamic variables around reference state
For completeness we have included the anelastic equations below. Any interested reader can find a much more detailed exposition of the anelastic approximation in Glatzmaier (1984) and references therein.

Barred symbols denote reference state quantities, all others perturbations.
{%\small
\begin{eqnarray}
\mathbf{\nabla}\cdot(\bar{\rho}\mathbf{v}) & = & 0\\
\bar{\rho}\frac{\partial \mathbf{v}}{\partial t} & = & -\mathbf{\nabla}\cdot(\bar{\rho}\mathbf{v}\otimes\mathbf{v})-\bar{\rho}\mathbf{\nabla}(\frac{P}{\bar{\rho}}+\bar{\Phi}_g)-\left(\frac{\partial \bar{\rho}}{\partial s}\right)_{P}s\bar{g}\mathbf{\hat{r}}+2\bar{\rho}\mathbf{v}\times\mathbf{\bar{\Omega}} \nonumber \\
& & +\mathbf{\nabla}\cdot\left(2\bar{\rho}\bar{\nu}(\bar{\mathbf{\mathsf{E}}}-\frac{1}{3}(\mathbf{\nabla}\cdot\mathbf{v})\bar{\mathbf{\mathsf{I}}})\right)\\
\bar{\rho}\bar{T}\frac{\partial s}{\partial t} & = & \mathbf{\nabla}\cdot(\bar{\kappa}\bar{\rho}\bar{T}\mathbf{\nabla}s)-\bar{T}\mathbf{\nabla}\cdot(\bar{\rho} s \mathbf{v})+\bar{\rho}(\bar{\epsilon}_{\mbox{\small nuc}}+\epsilon_{\mbox{\small nuc}}) \left[+ \bar{\rho}(\bar{\epsilon}_\nu+\epsilon_\nu)\right] \\
%+2\bar{\nu}\bar{\rho}\left(\mathbb{E}\cdot\mathbb{E}-\frac{1}{3}(\mathbf{\nabla}\cdot\mathbf{v})^2\right)
\rho & = & \left(\frac{\partial \bar{\rho}}{\partial s}\right)_P s +\left(\frac{\partial \bar{\rho}}{\partial P}\right)_s P
\end{eqnarray}}
Here $\mathbf{\mathsf{E}}$ is the rate of strain and $\mathbf{\mathsf{I}}$ the identity tensor. A vanishing coefficient of bulk viscosity is assumed for the viscous force calculation, and viscous heating is neglected. 

Of particular interest is the mass conservation equation in which the Eulerian time derivative of the density perturbation drops out. This in effect filters out sound waves and greatly ameliorates the Courant condition, which in turn allows larger timesteps. Note that by design anelastic codes cannot model any shocks, or (super)sonic flow in general. The anelastic approximation attempts to solve for an average solution of the flow, ignoring the propagation of any small scale pressure fluctuations.
  \subsection{Glatzmaier's Implementation}
%   - capability for rotation, magnetic fields
%   - constant, adiabatic reference state, depends only on radius
%   - fully spectral: Chebyshev in radial, spherical harmonics in angular
%   - cannot model center, always cut out a central sphere
%   - description of Gary's work with this code
The numerical implementation of these equations has been performed by Glatzmaier, and a detailed description of the code can be found in Glatzmaier (1984). The code is fully spectral, with an expansion in spherical harmonics to cover the angular variations and in Chebyshev polynomials for the radial dependence. 

The models described here have no radiative energy transport, so there are only two ways to transport energy: convective and diffusive heat flux. The nature of the diffusive heat flux is explained below in Section 2.5.
  \subsection{The Reference State}
One of the most important steps in producing these simulations is the generation of the reference state. We have incorporated a general equation of state and have added temperature and density dependent energy generation and neutrino losses.
    \subsubsection{Radial Structure}
%     - take output from 1D KEPLER code
%     - map rho,P onto polytrope => K,n
%     - solve Lane-Emden equation to get radial dependence: rho(r), P(r)
%     - take entropy
As the reference state is one-dimensional, we take advantage of our 1D stellar evolution code KEPLER (Weaver, Zimmerman, \& Woosley 1978) and fit its density and pressure to a polytrope ($P=K\rho^{1+1/n}$). For the radial dependence of these quantities we solve the resulting Lane-Emden equation. 
    \subsubsection{Equation of State}
%     - Timmes and Swesty Helmholtz E.O.S.
%     - had to ``invert the table'' to take rho,P,S as independent variables
%     - get T(r) from rho(r),P(r)
%     - also get various thermodynamic derivatives (list them)
The background temperature $\bar{T}$ as well as the necessary thermodynamic derivatives $\left(\frac{\partial \bar{\rho}}{\partial s}\right)_P,\left(\frac{\partial \bar{\rho}}{\partial P}\right)_s$ are determined using the publicly available Helmholtz code (Timmes \& Swesty 2000). 
    \subsubsection{Energy Generation and Neutrino Losses}
%     - fit KEPLER energy generation to eps_nuc=C*rho*T^n law (perturbation?)
%     - fit KEPLER neutrino losses to eps_nu=C*rho^m*T^n law
%     - use reference state T(r),rho(r) to calculate background egen and neutloss
It is important to capture the full temperature and density dependent nature of the energy generation within the convective region, and in the case of the oxygen shell also of the neutrino losses. For this reason we have added not only radially varying reference state nuclear energy generation and neutrino loss rates, but have also kept the perturbations to allow for differential heating between cold and hot convective bubbles.  The nuclear energy generation rate in the relevant zones is fit to the conventional $\epsilon_{\mbox{\small nuc}} \propto \rho T^n$ law, and the neutrino loss rate to $\epsilon_\nu \propto \rho^m T^n$.
%\begin{eqnarray}
%\bar{\epsilon}_{\mbox{nuc}} \propto \bar{\rho} \bar{T}^n & \rightarrow & \epsilon'_{\mbox{nuc}} \propto \left(\frac{\rho'}{\bar{\rho}} + n\frac{T'}{\bar{T}}\right) \\
%\bar{\epsilon}_\nu \propto \bar{\rho}^m \bar{T}^n & \rightarrow & \epsilon'_\nu \propto \left(m\frac{\rho'}{\bar{\rho}} + n\frac{T'}{\bar{T}}\right)
%\end{eqnarray}
  \subsection{Boundary Conditions}
%   - impermeable and stress free
%   - cannot model over/undershoot, limitation
%   - in future attach convectively stable regions to top and bottom
 The top and bottom boundaries are modeled as impermeable and stress-free ($v_\perp=0 \mbox{ and } \frac{\partial}{\partial r}\left(\frac{v_\parallel}{r}\right)=0$). This means that presently we are not able to study important phenomena such as convective overshoot. 

Generally we fix the entropy gradient at the bottom boundary by specifying a base luminosity, taken from KEPLER. For the boundary condition at the top we either force the entropy perturbation to vanish or we specify the top luminsity, i.e. the entropy gradient at the outer boundary.
  \subsection{Turbulent Diffusion}
%   - cannot resolve molecular level or small scale turbulence
%   - parametrize this by turbulent diffusion
%   - in these models taken as constant throughout star
%   - try to lower this artificial diffusion as much as possible
%    given a resolution.
%   - effective Reynolds and Rayleigh numbers depend on these turbulent
%     diffusion coefficients.
The diffusion of momentum and entropy by small-scale eddies is expected to dominate the diffusion due to radiative and molecular processes. The effects of all unresolved motions are parametrized as turbulent diffusion. In this model we have chosen constant turbulent viscous and thermal diffusivities, which we have tried to minimize to the extent permitted by the finite resolution. 
\section{Hydrogen Core Convection}
% - 15 solar mass star, halfway through main sequence
% - proof of principle, get familiar with code
% - not very violent, definitely anelastic material
% - one nonrotating, one rotating model
As our initial testcase for this code, in order to get familiar with it, we chose to model core convection of a $15 M_{\sun}$ star half-way through the main sequence. This stage of the star's life is not very violent; there are no neutrino losses, the energy generation doesn't have a very steep temperature dependence, there are no shocks, and density and pressure contrasts are expected to be small - a perfect case for an anelastic model. We view this model as a proof of principle, and as preparation for modeling more advanced stages of stellar evolution
  \subsection{Parameters}
The dependence of the reference state radius, density, temperature, and pressure on the Lagrangian mass coordinate are shown in Figure 1. Table 1 below summarizes the remaining parameters of interest.
\begin{table}
\caption{Parameters of the $15 M_{\sun}$ model of core convection during main sequence hydrogen burning.}
\center{
\begin{tabular}{cccccc}
\tableline
$M_{\mbox{\small base}}$ & $M_{\mbox{\small top}}$ & $L_{\mbox{\small base}}$ & $R_{\mbox{\small base}}$ & $R_{\mbox{\small top}}$  & $n_{\epsilon\mbox{\small ,nuc}}$\\
\tableline
$0.01 M_\odot$ & $4.16 M_\odot$ & $1000 L_\odot$ & $0.154 R_{\sun}$ & $1.18 R_{\sun}$ & $15.387$\\
\tableline
\tableline
\end{tabular}
}
\end{table}
$M_{\mbox{\small base}}$ is the mass of the central portion of the star that is cut out, and $M_{\mbox{\small top}}$ is the total mass within the convective core of this $15 M_{\sun}$ star. Even without the central region we are still modeling $99.8\%$ of the convective core, both by volume and mass.

\begin{figure}
\plottwo{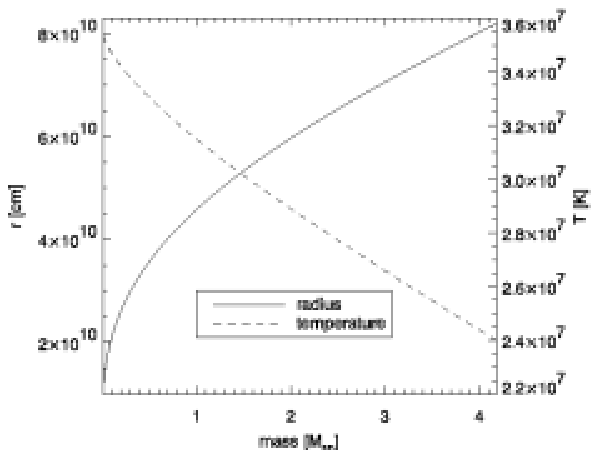}{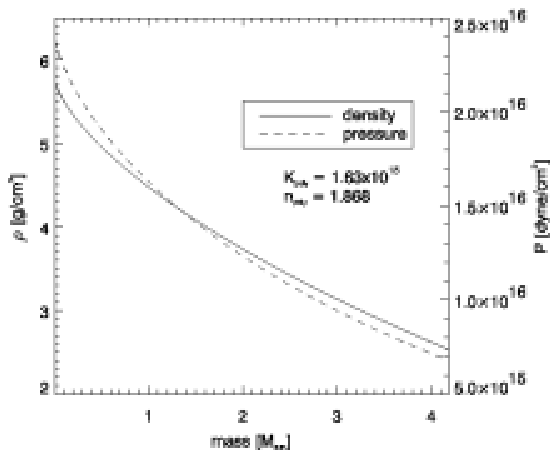}
\caption{15 $M_{\sun}$ hydrogen core convection model reference state.}
\end{figure}
  \subsection{Results}
% Results from three-dimensional models are not easily visualized. Here we have chosen to present equatorial and meridional slices, as well as Hammer-Aitoff equal area projections of constant radius surfaces. Also time-dependent information cannot be conveyed in a static plot, so please look at our website for additional plots and some movies. (http://www.supersci.org under Recent Results/Stellar Evolution)
\begin{table}
\caption{Numerical results of $15 M_{\sun}$ hydrogen core convection simulations.\\Resolution: $n_{\mbox{max}}=241, l_{\mbox{max}}=63, m_{\mbox{max}}=31$}
\center{
\begin{tabular}{cccccc}
\tableline
\vspace*{0.05in}
$\bar{\Omega}$ & $v_{\mbox{\small max}}$ & $<v>$ & $<\frac{1}{2}\rho v^2>$ & $N_{\mbox{\small Ra}}$ & $N_{\mbox{\small Re}}$ \\% & $\left(\frac{\delta\rho'}{\bar{\rho}}\right)_{\mbox{\small max}}$ \\ %& $\left(\frac{\delta P'}{\bar{P}}\right)_{\mbox{\small max}}$\\
\tableline
\vspace*{0.05in}
$0.0 \frac{\mbox{\small rad}}{\mbox{\small s}}$ & $1.5 \frac{\mbox{\small km}}{\mbox{\small s}}$ & $0.98 \frac{\mbox{\small km}}{\mbox{\small s}}$ & $2.0\times10^{10} \frac{\mbox{\small erg}}{\mbox{\small cm}^3}$ & $4\times 10^7$ & 1700 \\%& $10^{-5}$\\ % & $10^{-5}$\\
\tableline
\vspace*{0.05in}
$6.3\times10^{-5} \frac{\mbox{\small rad}}{\mbox{\small s}}$ & $3.5 \frac{\mbox{\small km}}{\mbox{\small s}}$ & $1.0 \frac{\mbox{\small km}}{\mbox{\small s}}$ & $2.5\times10^{10} \frac{\mbox{\small erg}}{\mbox{\small cm}^3}$ & $2\times 10^7$ & 3000 \\%& $2\times10^{-5}$\\ % & $2\times10^{-4}$\\
\tableline
\tableline
\end{tabular}
}
\end{table}
\begin{figure}[tp]
%\plotone{H.static.eps}
\plotfiddle{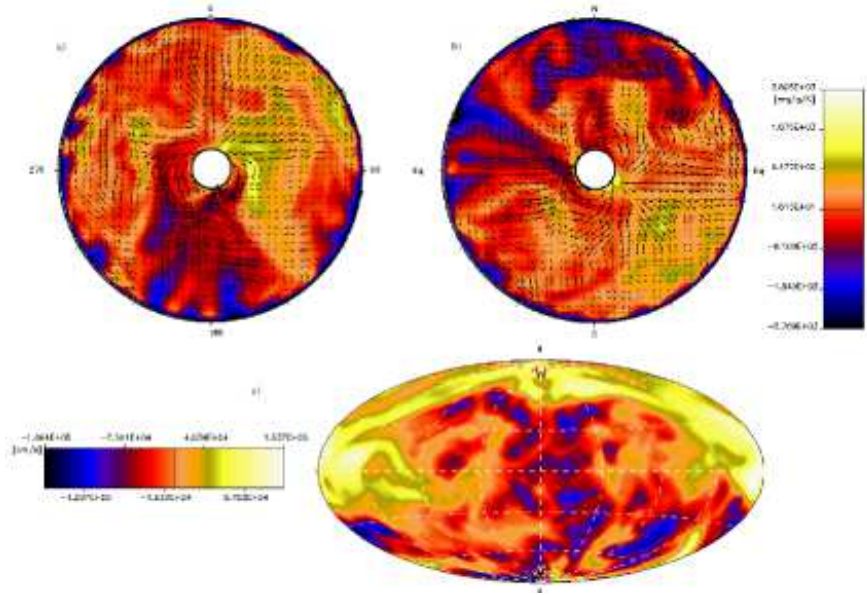}{2.8in}{0}{100}{100}{-175}{-15}
\caption{\small 15 $M_{\sun}$ hydrogen core convection model - non-rotating.\\
a) entropy perturbation in equatorial slice, b) entropy perturbation in meridional slice, c) radial component of velocity at constant radius surface ($68\%$, $R=5.6\times 10^{10} \mbox{cm}$) in equal-area projection.\\
a) and b) also show the planar velocity field}
\end{figure}

\begin{figure}[bp]
%\plotone{H.rotating.eps}
\plotfiddle{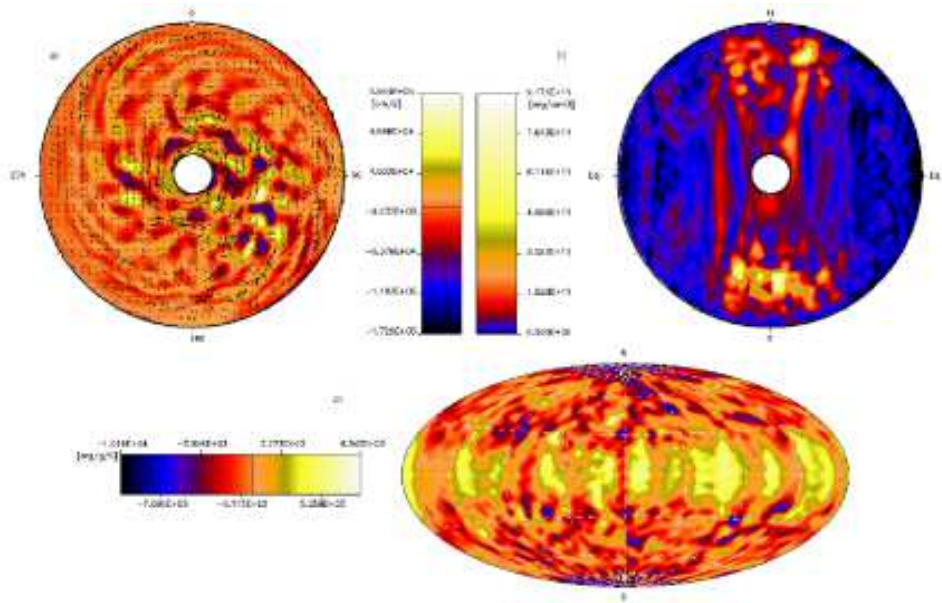}{2.8in}{0}{100}{100}{-175}{-15}
\caption{\small 15 $M_{\sun}$ hydrogen core convection model - rotating.\\
a) latitudinal component of velocity in equatorial slice with velocity field, b) kinetic energy in meridional slice, c) entropy perturbation at constant radius surface ($68\%$, $R=5.6\times 10^{10} \mbox{cm}$) in equal-area projection.}
\end{figure}

    \subsubsection{The Non-rotating Case}
The non-rotating model was run with a final resolution of 241 Chebyshev polynomials and spherical harmonics up to $l=63, m=31$. The model simulates about 44 days in the life of the star. An average convective speed of $\sim 10^5 \mbox{cm/s}$ implies a convective turnover time of $\tau_{\mbox{to}}=7\times 10^5 s \simeq 8 \mbox{days}$, and hence about 6 convective turnovers. The first row of Table 2 summarizes the numerical results of this simulation, and Figure 2 provides some visualization of the convective motions. Note that without rotation there is no preferred orientation of the model, which explains why figures 2a) and 2b) look so similar. Both slices seem to indicate a somewhat dipolar flow field, with high entropy gas rising on one side and falling at the other. Such dipolar flow pattern have been observed in fully compressible calculations as well (Jacobs, Porter, \& Woodward 1998).

All thermodynamic contrasts ($(\delta\rho'/\bar{\rho})_{\mbox{\small max}},(\delta P'/\bar{P})_{\mbox{\small max}},$ etc.) are of the order $10^{-5}$, comfortably below $1\%$. The mass-averaged reference state sound speed is about $700 \mbox{\small km/s}$, so the flow is completely subsonic. We conclude that this problem is very suited for an anelastic treatment, as expected.
\subsubsection{The Rotating Case}
The rotating model was run for about 120 days in the star's life, with the same resolution as the non-rotating case. This amount of star time corresponds to about 15 convective turnovers. As is immediately obvious by looking at Figure 3, rotation has a significant effect on the convective flow pattern. The rotational axis breaks the symmetry of the problem, so the flow behaves differently in the equatorial plane than in any meridional plane and the dipolar flow pattern is no longer visible. In the meridional slice the strong, high kinetic energy flows are concentrated close to the axis, with convection being strongly suppressed in the equatorial regions further away from the axis. We also see evidence of Taylor columns, elongated, columnar flow structures parallel to the rotation axis. 
%\pagebreak
%\pagebreak
\section{Oxygen Shell Convection}
Of the advanced stages of stellar evolution the oxygen shell burning phase is of particular importance. For one, convection during this brief time before core collapse might produce an asymmetry, leading to non-uniform mixing and perhaps even an asymmetric supernova explosion. Several two-dimensional studies of this problem have been conducted (Arnett 1994; Bazan \& Arnett 1998; Asida \& Arnett 2000), and several of these simulations found close to sonic flow, large density and pressure contrasts, and hence significant acoustic fluxes. 

As an alternative to these simulations, we present results of our anelastic model of a similar problem - oxygen shell convection in a $25 M_{\sun}$ star, shortly after core oxygen depletion. As shown below, we have found a solution within the anelastic approximation, with small density and pressure contrasts, and subsonic flows.
  \subsection{Parameters}
\begin{table}
\caption{Parameters of the $25 M_{\sun}$ model of convection during oxygen shell burning, about 4 days before core collapse.}
\center{
\begin{tabular}{cccccc}
\tableline
$M_{\mbox{\small base}}$ & $M_{\mbox{\small top}}$ & $L_{\mbox{\small base}}$ & $R_{\mbox{\small base}}$ & $R_{\mbox{\small top}}$\\ %  & $n_{\epsilon\mbox{\small ,nuc}}$\\
\tableline
$1.2 M_\odot$ & $1.9 M_\odot$ & $19500 L_\odot$ & $0.0064 R_{\sun}$ & $0.012 R_{\sun}$\\% & $15.387$\\
\tableline
\tableline
\end{tabular}
}
\end{table}
During the modeled phase the rate of nuclear energy generation exceeds the rate of neutrino losses by about a factor two. The net energy generation rate is positive, $\int_{\mbox{\tiny base}}^{\mbox{\tiny top}} 4 \pi r^2 \rho(r) (\bar{\epsilon}_{\mbox{\small nuc}}(r) + \bar{\epsilon}_\nu(r)) = 1.5\times10^{45} \mbox{erg/s}$. KEPLER indicates that this energy is going into $PdV$ work and into change of the gravitational potential. In the models presented here we have chosen to allow this excess energy to escape off the top, as we are currently unable to model expansion.
\begin{figure}
\plottwo{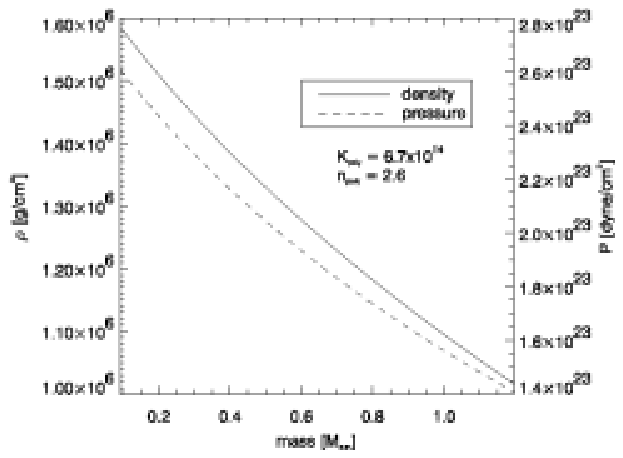}{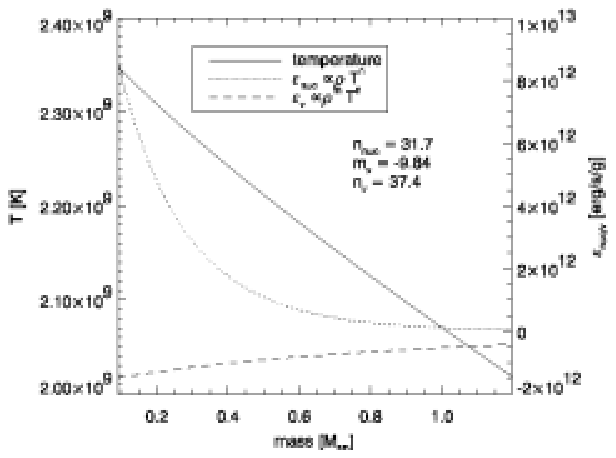}
\caption{25 $M_{\sun}$ oxygen shell convection model reference state.}
\end{figure}
  \subsection{Results}
\begin{figure}[tp]
%\plotone{O.static.eps}
\plotfiddle{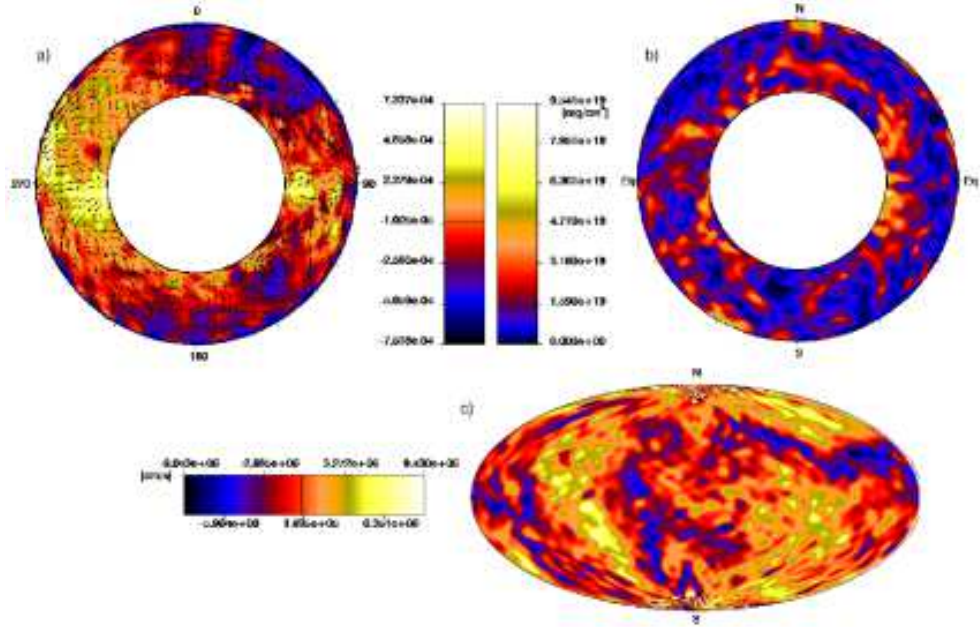}{3.0in}{0}{100}{100}{-175}{-15}
\caption{\small 25 $M_{\sun}$ oxygen shell convection model - non-rotating.\\
a) entropy perturbation in equatorial slice with velocity field, b) kinetic energy in meridional slice, c) radial component of velocity at constant radius surface ($84\%$, $R=7.4\times 10^8 \mbox{cm}$) in equal-area projection.}
\end{figure}

\begin{figure}[bp]
%\plotone{O.rotating.eps}
\plotfiddle{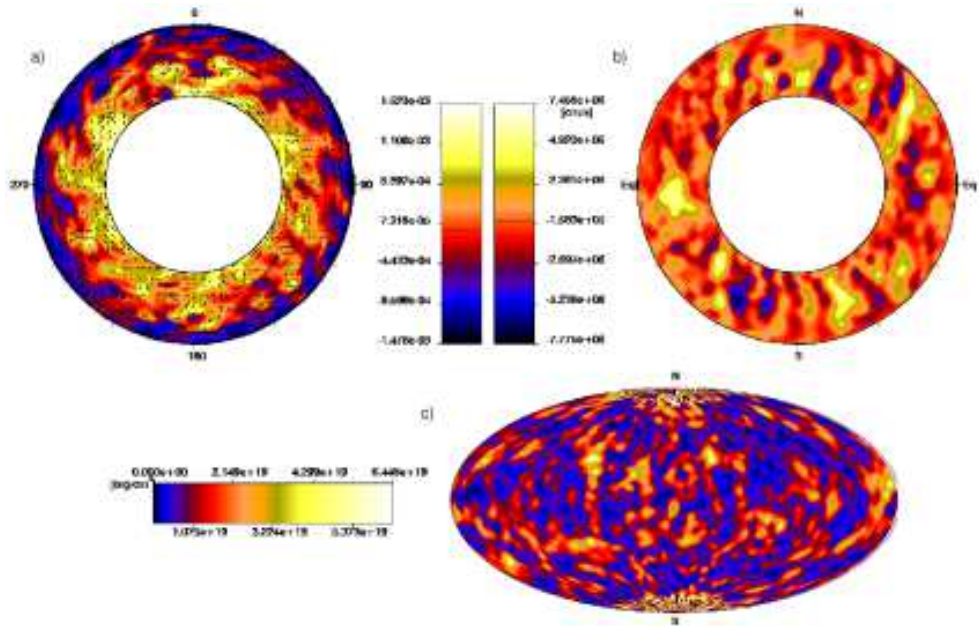}{3.0in}{0}{100}{100}{-175}{-15}
\caption{\small 25 $M_{\sun}$ oxygen shell convection model - rotating.\\
a) entropy perturbation in equatorial slice with velocity field, b) radial component of velocity in meridional slice, c) kinetic energy at constant radius surface ($68\%$, $R=6.8\times 10^8 \mbox{cm}$) in equal-area projection.}
\end{figure}
\begin{table}
\caption{Numerical results of $25 M_{\sun}$ oxygen shell convection simulations.\\Resolution: $n_{\mbox{max}}=145, l_{\mbox{max}}=63, m_{\mbox{max}}=31$}
\center{
\begin{tabular}{cccccc}
\tableline
\vspace*{0.05in}
$\bar{\Omega}$ & $v_{\mbox{\small max}}$ & $<v>$ & $<\frac{1}{2}\rho v^2>$ & $N_{\mbox{\small Ra}}$ & $N_{\mbox{\small Re}}$\\ % & $\left(\frac{\delta\rho'}{\bar{\rho}}\right)_{\mbox{\small max}}$ \\ %& $\left(\frac{\delta P'}{\bar{P}}\right)_{\mbox{\small max}}$\\
\tableline
\vspace*{0.05in}
$0.0 \frac{\mbox{\small rad}}{\mbox{\small s}}$ & $180 \frac{\mbox{\small km}}{\mbox{\small s}}$ & $49 \frac{\mbox{\small km}}{\mbox{\small s}}$ & $1.7\times10^{19} \frac{\mbox{\small erg}}{\mbox{\small cm}^3}$ & $5\times 10^7$ & 3000\\ % & $10^{-5}$\\ % & $10^{-5}$\\
\tableline
\vspace*{0.05in}
$0.08 \frac{\mbox{\small rad}}{\mbox{\small s}}$ & $170 \frac{\mbox{\small km}}{\mbox{\small s}}$ & $40 \frac{\mbox{\small km}}{\mbox{\small s}}$ & $1.2\times10^{19} \frac{\mbox{\small erg}}{\mbox{\small cm}^3}$ & $6\times 10^7$ & 2500\\ % & $2\times10^{-5}$\\ % & $2\times10^{-4}$\\
\tableline
\tableline
\end{tabular}
}
\end{table}
\begin{figure}
\plottwo{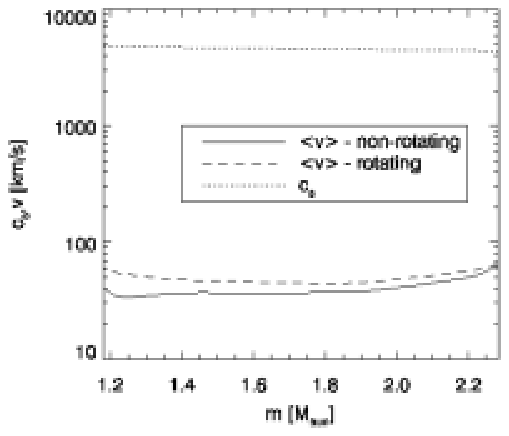}{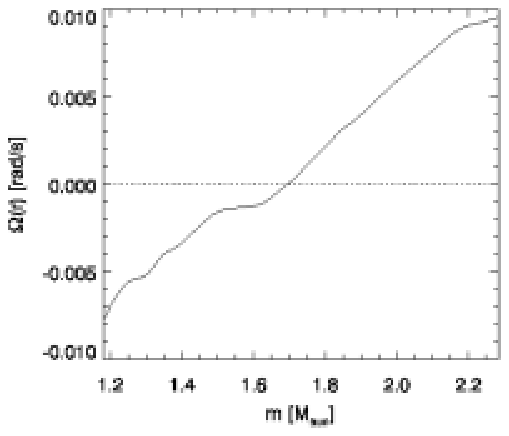}
\caption{25 $M_{\sun}$ oxygen shell convection model. Left: Comparison of convective velocities with sound speed. Right: Differential rotation in equatorial plane of the rotating model.}
\end{figure}
\begin{table}
\caption{Thermodynamic contrasts for the $25 M_{\sun}$ oxygen shell convection simulations.}
\center{
\begin{tabular}{ccccc}
\tableline
\vspace*{0.05in}
& $\left(\frac{\delta\rho'}{\bar{\rho}}\right)_{\mbox{\small max}}$ & $\left(\frac{\delta s'}{\bar{s}}\right)_{\mbox{\small max}}$ & $\left(\frac{\delta T'}{\bar{T}}\right)_{\mbox{\small max}}$ &$\left(\frac{\delta P'}{\bar{P}}\right)_{\mbox{\small max}}$\\
\tableline
\vspace*{0.05in}
non-rotating & $2 \times 10^{-3}$ & $1 \times 10^{-3}$ & $6\times 10^{-4}$ & $7 \times 10^{-4}$ \\
\tableline
\vspace*{0.05in}
rotating & $3 \times 10^{-3}$ & $3 \times 10^{-3}$ & $1\times 10^{-3}$ & $1 \times 10^{-3}$ \\
\tableline
\tableline
\end{tabular}
}
\end{table}
    \subsubsection{The Non-rotating Case}
The non-rotating model was run for an equivalent star time of $6500$ seconds. Given an average speed of about $50$ km/s we determine the convective turnover time to be about $70$ seconds, which implies that we followed convection for about $90$ turnovers. Figure 5 shows that the convective speed remains around two orders of magnitude below the sound speed throughout the entire modeled region. All thermodynamic contrasts are below the $1\%$ level, as shown in Table 5. It is important to note that nothing in the model forces the resulting convective velocities to be subsonic and nothing prevents the thermodynamic perturbations from rising to a level comparable to the background. The fact that this solution remained subsonic and with small thermodynamic contrasts implies to us that the anelastic approximation is a valid one for the problem of the convective oxygen shell.

The entropy perturbations in the equatorial plane (see Figure 6) indicate a flow pattern dominated by four large cells. High entropy fluid is rising towards $90\deg$ and $270\deg$, and low entropy fluid is falling along the $0\deg$ and $180\deg$ directions. We have not yet determined how much of this simple structure is dependent on Rayleigh and Reynolds numbers. It is possible that higher resolution simulations, with higher Rayleigh and Reynolds numbers, will destroy this pattern.
\subsubsection{The Rotating Case}
The rotating case was run for $3300$ seconds, which corresponds to about 50 convective turnovers. Again the convective speed is very subsonic and the thermodynamic contrasts are still below $1\%$, albeit somewhat higher than in the non-rotating case.

Although the reference state is rotating with a fixed angular speed ($\bar{\Omega}=0.08$ {\small rad/s}) the flow is free to arrange itself into any form of differential rotation. The result can be seen in Figure 5. Our rotating model shows a lot of differential rotation, with the outer regions of the equatorial plane rotating nearly twice as fast as the inner ones. Just like in the hydrogen core convection model we see Taylor-column-like structures parallel to the rotation axis in the meridional planes. 
\vspace*{-0.2in}
\section{Conclusions}
\vspace*{-0.1in}
Anelastic hydrodynamics appears to be a useful tool for studying convection in the interiors of massive stars in three dimensions. As long as the problem at hand satisfies the anelastic conditions of subsonic flow speeds and small thermodynamic perturbations, a lot of progress can be made with an anelastic code in a shorter amount of time and with less computational resources than with a fully compressible code.

In this paper we have presented results of our efforts to model convection in the core of $15 M_{\sun}$ main sequence star as well as in the oxygen burning shell of a $25 M_{\sun}$ star at a much more advanced stage of stellar evolution. The resulting convective speeds are in agreement with results from KEPLER using mixing length theory. The resulting three dimensional flow pattern are not always uniform, and can show strong asymmetries, such as the dipolar flow in the hydrogen core convection model. We have found a significant difference in the flow pattern of our rotating models compared to the non-rotating ones. The presence of rotation breaks the spherical symmetry and allows more vigorous convection along the axis of rotation. The rotating oxygen shell model showed a significant degree of differential rotation.

This work was supported by the Scientific Discovery though Advanced Computing (SciDAC) Program of the DOE (DE-FC02-01ER41176).

\end{document}